\documentclass[11pt]{article}
\usepackage{amssymb,amsmath} 
\usepackage{graphicx}
\usepackage[colorlinks=true, pdfstartview=FitV, linkcolor=blue, citecolor=blue, urlcolor=blue]{hyperref}

\def\div{\nabla\cdot } 
\def\pl{\partial}

\newcommand{\rt}{(\3r,t)}

\newcommand{\bib}{\bibitem}

\newcommand{\nt}{\notag}
\newcommand{\ci}{\cite}
\newcommand{\eq}{\eqref}

\newcommand{\bx}[1]{\boxed{\ #1\ }}

\newcommand{\lp}{\left(}
\newcommand{\rp}{ \right)}
\newcommand{\lb}{ \left[}
\newcommand{\rb}{\right]}

\newcommand{\harr}[1]{\smash{\mathop{\hbox to .5in{\ \rightarrowfill\ }}
      \limits^{#1}}}

\newcommand{\0}[1]{{(#1)}}

\newcommand{\2}[1]{{\tilde #1}}
\newcommand{\3}[1]{{\boldsymbol #1}}

\newcommand{\bh}[1]{\mbox{\boldmath{{$\rm\hat#1$}}}}
\newcommand{\bt}[1]{\mbox{\boldmath{{$\tilde#1$}}}}

\newcommand{\ctil}[1]{{\mathcal{\tilde{ #1}}}}

\def\a{\alpha} 
\def\b{\beta} 

\def\c{\chi}
\def\d{\delta} 
\def\e{\varepsilon} 

\def\f{\phi} 
 
\def\g{\gamma}
\def\h{\eta} 

\def\k{\kappa}

\def\m{\mu} 

\def\o{\omega} 
\def\p{\pi} 

\def\q{\theta} 
\def\vq{\vartheta} 
\def\r{\rho}
\def\vr{\varrho} 
\def\s{{\sigma}} 

\def\t{\tau} 
 
\def\x{\xi}
\def\y{\psi} 
\def\z{\zeta}

\newcommand{\db}{{\,{\rm d}\kern-1.6ex-}}
\newcommand{\dir}{{\pl\kern-1.2ex {/}}}
\newcommand{\dd}{{\rm d}}

\newcommand{\app}{\approx} 
\newcommand{\cc}[1]{{{\mathbb C\hskip.5pt}^{#1}}}

\newcommand{\curl}{\nabla\times} 

\newcommand{\eg}{{\it e.g., }}

\newcommand{\grad}{\nabla}

\newcommand{\ie}{{\it i.e., }}

\def\iff{\ \Leftrightarrow\ }
\newcommand{\im}{{\,\rm Im}\ }  
\newcommand{\imp}{\ \Rightarrow\ }

\newcommand{\ir}{\int_{-\infty}^\infty}  

\newcommand{\lra}{\leftrightarrow}

\newcommand{\plra}{\pl^{\kern-1.25ex^\lra}}
\newcommand{\qq}{\quad} 
\newcommand{\qqq}{\qquad} 
\newcommand{\re}{{\,\rm Re}\  }   

\newcommand{\rr}[1]{{{\mathbb R}^{#1}}}

\newcommand{\sgn}{{\,\rm Sgn \,}}
\newcommand{\sh}[1]{\hskip#1ex} 
\newcommand{\sr}{\sqrt}

\newcommand{\orr}{{(\3r)}}

\def\Xint#1{\mathchoice
   {\XXint\displaystyle\textstyle{#1}}%
   {\XXint\textstyle\scriptstyle{#1}}%
   {\XXint\scriptstyle\scriptscriptstyle{#1}}%
   {\XXint\scriptscriptstyle\scriptscriptstyle{#1}}%
   \!\int}
\def\XXint#1#2#3{{\setbox0=\hbox{$#1{#2#3}{\int}$}
     \vcenter{\hbox{$#2#3$}}\kern-.5\wd0}}

\def\ppint{\Xint-}

\def\VE{\vfill\eject}

\def\bib#1{\bibitem[#1]{#1}}

\begin{document}

\title{Electromagnetic helicity wavelets:\\ a model for quasar engines?}

\author{Gerald Kaiser\\
\href{http://wavelets.com}{Center for Signals and Waves}\\ Austin, TX\\
kaiser@wavelets.com
}

\maketitle

\begin{abstract}\noindent 
The complex distance function $\z$, which plays a prominent role in the definition of scalar (acoustic) wavelets, is found to determine a complex extension of the spherical coordinate system that is ideally suited for the construction of highly focused electromagnetic beams with helicities conforming to the oblate spheroidal geometry of $\z$. This is  used to build a basis of electromagnetic wavelets $\3F^m=\3E^m+i\3B^m$ radiated or absorbed by the branch disk $\5D$ of $\z$. $\3F^m$ has integer angular momentum $m$ around the axis of $\5D$ and definite spheroidal helicity. We use a regularization method to compute its singular charge-current density and show that the total charge vanishes. Hence $\3F^m$ is due solely to electric and magnetic polarization currents. $\5D$ acts as a magnetic dipole antenna, and its axis as a coupled electric dipole antenna. We propose this as an idealized electromagnetic model for quasars (without gravity, \ie in flat spacetime), with $\5D$ representing the accretion disk and the vortex singularities along its axis representing the jets. In the regularized version, the accretion disk is represented by a solid, flat oblate spheroid and the jets by two solid, narrow semi-hyperboloids, as shown in Figure \ref{F:EMDipole}.
\end{abstract}

\VE

\tableofcontents

\section{The complex helicity basis}

\begin{figure}[ht]
\begin{center}
\includegraphics[width=2.5 in]{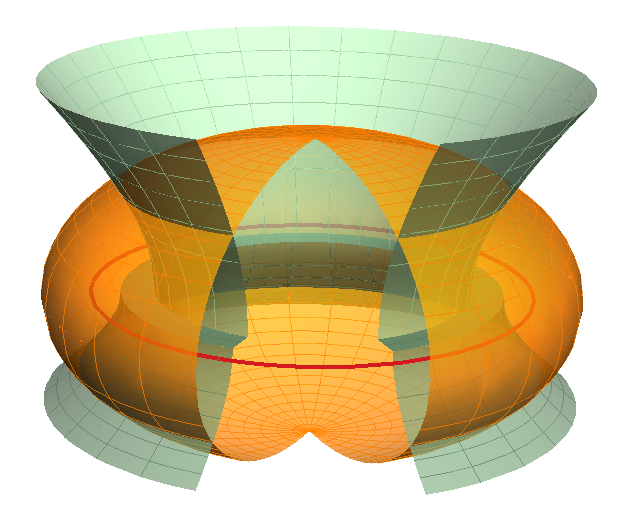}
\caption{\small The real and imaginary parts of $\z=\x-i\h$ form an oblate spheroidal coordinate system with the $z$ axis along $\3a$. Shown are cut-away views of an oblate spheroid $\5E$ with $\x=\e\, (\e=0.7a)$, a semi-hyperboloid $\5H_+$ with $\h=b\, (b=0.8a)$, and a semi-hyperboloid $\5H_-$ with $\h=-b' \,(b'=0.5a)$. Also shown is the branch circle $\5C=\pl\5D$, which is spanned by  the branch disk $\5D$.
$\5C$ is the common \sl focal circle \rm of all the spheroids and hyperboloids. As $\e\to 0$, $\5E$ shrinks to a double cover of $\5D$.  As $b\to a$, $\5H_\pm$ shrink to the positive and negative $z$ axis. According to \eq{far}, the spheroids $\5E$ are deformations of the spheres $r=$ constant and the semi-hyperboloids $\5H_\pm$ are deformations of the cones $a\cos\q=$ constant, whose vertex $\3r=\30$ expands to the circle $\{\3r: \r=\sr{a^2-\h^2},\, z=0\}$.}
\label{F:Fig_OS}
\end{center}
\end{figure}

A rich geometry ensues when the Euclidean distance in $\rr n$ is continued analytically \ci{K0}.
Here we specialize to $\rr 3$ and use the notation introduced in \ci{HK9}. Define the complexification $\z$ of the radial coordinate $r$ by
\begin{gather}\label{zeta}
\z=\sr{(\3r-i\3a)^2}=\sr{\r^2+\2z^2}\equiv \x-i\h,\ \ \3r=(x,y,z)\in\rr3,\\
\3a=a\bh z,\qq \r=\sr{x^2+y^2},\qq  \2z=z-ia.\nt
\end{gather}
The singularity of $r$ at the origin expands to \sl the branch circle \rm of $\z$,
\begin{align*}
\5C\equiv \{\3r: \z=0\}=\{(x,y,z): x^2+y^2\equiv  a^2,\ z=0\},
\end{align*}
and continuation of $\z$ along any simple closed loop linking with $\5C$ gives $-\z$. Hence $\z$ is double-valued. To make it single-valued, it is necessary to close $\5C$ with any \sl membrane \rm whose boundary is $\5C$, the simplest of which is the disk
\begin{align*}
\5D=\{(x,y,z): x^2+y^2\le a^2,\ z=0\}=\{\3r: \x=0\}.
\end{align*}
$\5D$ is a branch cut of $\z$, and we choose the branch with $\x\ge 0$ which reduces to $+r$ when $a\to0$. On crossing $\5D$, $\h$ changes sign.  As shown in the Appendix \eq{zeta2}, $(\x,\h)$ are related to the cylindrical coordinates $(\r,z)$ by
\begin{align*}
az=\x\h,\qqq a^2\r^2=(a^2+\x^2)(a^2-\h^2),
\end{align*}
hence $-a\le\h\le a$. Far from $\5D$, we have
\begin{align}\label{far}
r\gg a\imp \x\app r\ \ \hbox{and}\ \ \h\app\bh a\cdot\3r =a\cos\q.
\end{align}
The level sets of $\x$ and $\h$ define an \sl oblate spheroidal coordinate system, \rm as explained in Figure \ref{F:Fig_OS} (see \ci{HK9} for details).

We now use $\z$ to defined a complexified version $\vq$ of the polar coordinate $\q$:
\begin{align}\label{vq}
\sin\vq=\frac\r\z,\qq \cos\vq=\frac{\2z}\z.
\end{align}
Like $\z$, $\vq$ is singular on $\5C$ and discontinuous on the interior of $\5D$. By applying the gradient operator with respect to $\3r$ to the coordinates $(\z,\vq,\f)$, we obtain vectors $(\bh\z, \bh\vq, \bh\f)$ extending the spherical basis $(\bh r, \bh\q, \bh\f)$:
\begin{align*}
\bh\z&\equiv \grad\z=\frac{\3r-i\3a}\z=\frac{\bh\r\r+\bh z\2z}\z=\bh\r\sin\vq+\bh z\cos\vq\\ 
\grad\vq&=\frac{\bh\vq}\z\ \ \hbox{where}\ \ \bh\vq=\bh\r\cos\vq-\bh z\sin\vq\\
\grad\f&=\frac{\bh\f}\r=\frac{\bh\f}{\z\sin\vq}.
\end{align*}
These vectors form a \sl complex-orthogonal, \rm oriented basis in $\cc3$, \ie
\begin{gather*}
\bh\z^2=\bh\vq^2=\bh\f^2=1,\qq
\bh\z\cdot\bh\vq=\bh\vq\cdot\bh\f=\bh\f\cdot\bh\z=0\\
\bh\z\times\bh\vq=\bh\f,\qq \bh\vq\times\bh\f=\bh\z, \qq \bh\f\times\bh\z=\bh\vq.
\end{gather*}
In this basis, the gradient operator is 
\begin{align*}
\grad=\bh\z\pl_\z+\frac{\bh\vq}\z\pl_\vq+\frac{\bh\f}{\z\sin\vq}\pl_\f.
\end{align*}
We now define a related basis which will be especially useful for studying electromagnetic radiation. Let
\begin{align*}
\k=\ln\tan(\vq/2)=\frac12\ln\frac{1-\cos\vq}{1+\cos\vq}=\frac12\ln\frac{\z-\2z}{\z+\2z},
\end{align*}
so that
\begin{align}\label{kappa}
\frac{\dd\k}{\dd\vq}=\frac1{\sin\vq}.
\end{align}
Let
\begin{align*}
\y_\pm\equiv \f\pm i\k,\qqq  \pl_\pm\equiv \frac{\pl}{\pl\y_\pm}=\frac12\lp\pl_\f\mp i\pl_\k\rp
\end{align*}
and define the vectors
\begin{align}\label{cpm0}
\3\c_\pm\equiv \grad\y_\pm=\frac{\bh\f}\r\pm\frac i{\sin\vq}\frac{\bh\vq}\z=\frac{\bh\f\pm i\bh\vq}\r.
\end{align}
Note that the factor $\dd\k/\dd\vq$ transforms the singularity $1/\z$ of $\grad\vq$ into the singularity $1/\r$ of $\grad\f$, so that $\bh\f$ and $\bh\vq$ can be combined into $\bh\f\pm i\bh\vq$. This combination will be the key to helicity in our system, and $(\bh\z,\3\c_+,\3\c_-)$ will be called the \sl helicity basis. \rm

Applying the chain rule in the coordinate system $(\z,\y_+, \y_-)$ gives the gradient operator in the helicity basis:
\begin{align*}
\grad=\bh\z\pl_\z+\3\c_+\pl_++\3\c_-\pl_-\,.
\end{align*}
Note that
\begin{align}\label{cpm}
&\bh\z\cdot\3\c_\pm=\3\c_\pm^2=0,&& \3\c_+\cdot\3\c_-=\frac2{\r^2}\\
&\bh\z\times\3\c_\pm=\pm i\3\c_\pm,&& \3\c_+\times\3\c_-=\frac{2i\bh\z}{\r^2},\nt
\end{align}
so $\3\c_\pm$ are a pair of \sl null, transversal eigenvectors \rm of the \sl operator \rm $\bh\z\times$ with eigenvalues $\pm i$. We shall see that the eigenvalue $i$ corresponds to \sl advanced (absorbed) solutions \rm and  $-i$ corresponds to \sl retarded (emitted) solutions. \rm 

In addition to the complex angular coordinates $\y_\pm$, define the \sl complex retarded and advanced time coordinates \rm
\begin{align*}
u=\t-\z,\qq v=\t+\z,\ \ \hbox{where}\ \ \t=t-is
\end{align*}
is complexified time \ci{K11} and we have chosen units in which the speed of light $c=1$. The functions $(u,v, \y_+, \y_-)$ form a \sl  null coordinate system \rm in spacetime related to the Newman-Penrose formalism. 

Note that $(u,v)$ are singular on $\5D$ and $\y_\pm$ are singular on the $z$ axis \eq{cpm0}. Hence our coordinate system is singular on the set
\begin{align}\label{5Q}
\5Q=\{\3r:\x=0\}\cup\{\3r: \r=0\}=\5D\cup\5Z_+\cup\5Z_-,
\end{align}
where $\5Z_+$ and $\5Z_-$ are the positive and negative $z$ axis, respectively.
We shall see that these coordinates are particularly well suited for representing electromagnetic fields radiated or absorbed by $\5D$. The coordinate singularities will be natural for describing sources on $\5Q$, just the singularity of the spherical coordinate system $(r,\q,\f)$ at the origin is natural for describing point sources at $r=0$.

\section{EM fields in the helicity basis}\label{S:general}

We consider the class of real electromagnetic fields $(\3E, \3B)$ represented in the complex coordinates $(u,v,\y_\pm)$ by
\begin{align}\label{F}
\3E+i\3B\equiv \3F=\a\3C+\b\3\c_++\g\3\c_-,
\end{align}
where
\begin{align*}
\3C=-\grad\frac2\z=\frac{2\bh\z}{\z^2}
\end{align*}
is the complexified Coulomb field \ci{N73} and $\a,\b,\g$ are analytic functions of $(u,v,\y_\pm)$ when $\3r\notin\5Q$. The advantage of the representation \eq{F} is that
\begin{align*}
\curl\3C=\curl\3\c_\pm=\30,
\end{align*}
which will make it easy to compute $\curl\3F$, and 
\begin{align}\label{raw}
\3r\notin\5Q\imp \div\3C= \div\3\c_\pm =0,
\end{align}
which will make it easy to compute $\div\3F$ outside of $\5Q$.

The charge-current density $(\vr,\3J)$ of $\3F$ is \sl defined\,\rm\footnote{We use Lorentz-Heaviside units, where $\e_0=\m_0=1$, and set $c=1$. The real parts of \eq{Max} give the inhomogeneous Maxwell equations with $(\re\vr, \re\3J)$ as the electric charge-current density, and imaginary parts state that  $(\im\vr, \im\3J)$ is the \sl magnetic \rm charge-current density. To avoid magnetic monopoles, we will show that the total magnetic charge vanishes, so  $(\im\vr, \im\3J)$ is due solely to magnetic polarization. 
}
\ci{K3}  by
\begin{align}\label{Max}
\vr\equiv \div\3F\ \ \hbox{and}\ \  -\3J\equiv \pl_t\3F+i\curl\3F,
\end{align}
where the derivatives must be interpreted in a \sl distributional \rm sense and will be seen to be distributions (generalized functions) supported on $\5Q$. We shall compute these sources by \sl regularizing \rm the field $\3F$ and then taking the limit as the regularization is removed. But first, let us find the conditions on $(\a,\b,\g)$ which make $\3F$ a solution of the \sl homogeneous \rm Maxwell equations outside the singularity set $\5Q$. 

Note that 
\begin{align}\label{pltz}
\pl_t\a=\a_u+\a_v\ \ \hbox{and}\ \ \pl_\z\a=\a_v-\a_u,
\end{align}
where the subscripts denote partial derivatives, and the same goes for $\b,\g$. Taking the divergence of \eq{F} and using the properties of $\3C$ and $\3\c_\pm$ gives
\begin{align}\label{rho1}
\vr=2\frac{\a_v-\a_u}{\z^2}+2\frac{\b_-+\g_+}{\r^2}.
\end{align}
Furthermore,
\begin{align*}
\pl_t\3F&=2\frac{\a_u+\a_v}{\z^2}\bh\z+(\b_u+\b_v)\3\c_++(\g_u+\g_v)\3\c_-\\
i\grad\a\times\3C&=i(\a_+\3\c_++\a_-\3\c_-)\times\frac{2\bh\z}{\z^2}
=\frac{2\a_+}{\z^2}\3\c_+-\frac{2\a_-}{\z^2}\3\c_-\\
i\grad\b\times\3\c_+&=i(\b_v-\b_u)\bh\z\times\3\c_++i\b_-\3\c_-\times\3\c_+
=(\b_u-\b_v)\3\c_++\frac{2\b_-}{\r^2}\bh\z\\
i\grad\g\times\3\c_-&=i(\g_v-\g_u)\bh\z\times\3\c_-+i\g_+\3\c_+\times\3\c_-
=(\g_v-\g_u)\3\c_--\frac{2\g_+}{\r^2}\bh\z.
\end{align*}
Collecting terms, we  have
\begin{align}\label{J1}
\sh{-3}-\3J&=2\bh\z\lb\frac{\a_u+\a_v}{\z^2}+\frac{\b_--\g_+}{\r^2}\rb
+2\3\c_+\lb\b_u+\frac{\a_+}{\z^2}\rb+2\3\c_-\lb\g_v-\frac{\a_-}{\z^2}\rb.
\end{align}
Requiring the sources \eq{rho1} and \eq{J1} to vanish outside the singularity set $\5Q$, 
the \sl homogeneous \rm Maxwell equations are thus equivalent to
\begin{align*}
(\a_v-\a_u)\sin^2\vq+\b_-+\g_+=0\,\\
(\a_u+\a_v)\sin^2\vq+\b_--\g_+=0\,\\
\a_+-\z^2\b_u=0\,\\
\a_-+\z^2\g_v=0.
\end{align*}
By simplifying and using
\begin{align*}
\sin\vq=\frac1{\cosh\y},\qq \cos\vq=\tanh\y,\qq \y=\frac{\y_++\y_-}2,\qq \z^2=\frac{(u-v)^2}4,
\end{align*}
this can be expressed in terms of of the coordinates $(u,v,\y_\pm)$ as 
\begin{align}\label{rels}
\sh{-2} \bx{\a_u=\g_+\cosh^2\y, \ \  \a_v=-\b_-\cosh^2\y,\ \ 
\a_+=\z^2\b_u, \ \  \a_-=-\z^2\g_v.}
\end{align}
These equations, which relate the longitudinal component $\a$ of the field to its transversal components $\b,\g$, express the homogeneous Maxwell equations in the coordinate system $(u,v,\y_\pm)$.

Newman's holomorphic Coulomb field, representing the electromagnetic part of a spinning, charged (Kerr-Newman) black hole (\ie ignoring gravity) \ci{N73}, is obtained by choosing 
\begin{align*}
\a=\frac12,\qq \b=\g=0\imp\3F=\frac{\bh\z}{\z^2}=-\grad\frac1\z.
\end{align*}
As proved in \ci{K4}, the charge-current density in this case is that of a disk $\5D$ spinning rigidly at the extreme relativistic angular velocity $\o=c/a$, so that the boundary $\5C=\pl\5D$ spins at the speed of light. This kind of singular behavior is expected of black holes.

The \sl coherent electromagnetic wavelets \rm constructed in \ci{K11},
\begin{align}\label{F-}
\3F=g'(\t-\z)\3\c_-\,,
\end{align}
are obtained by choosing
\begin{align*}
\a=\b=0,\qq\g=g'\0u,
\end{align*} 
which solves \eq{rels}.
Again, these are extreme solutions of Maxwell's equations whose energy propagates at the speed of light everywhere outside the singularity set $\5Q$; see \ci{K11b, K12} for a detailed discussion.

The function $g'$ in \eq{F-} is the complex derivative of the \sl analytic signal \rm $g$ of an arbitrary pulse function $g_0\0t$, defined by
\begin{align}\label{ast}
g\0\t=\frac1{2\p i}\ir\frac{g_0(t')\,\dd t'}{\t-t'},\qq \t=t-is,\qq s\ne 0.
\end{align}
This is a special case of the \sl multidimensional analytic-signal transform \rm  \ci{K3, K11a}.
If $g_0\0t$ is at all reasonable (\eg square-integrable), then $g(t-is)$ is analytic outside the real axis and its real and imaginary parts are smeared version (to scale $|s|$) of $g_0\0t$ and its Hilbert transform, respectively. The farther we get from the real time axis (by increasing $|s|$), the more smeared out and weaker the analytic signal $g(t-is)$ becomes, and the same goes for its complex derivative $g'(t-is)$. Since
\begin{align*}
g'(\t-\z)=g'(t-\x-i(s-\h)),
\end{align*}
it follows that $g'(\t-\z)$ is analytic for $s>a$ (since then $s-\h\ge s-a>0$ for all $\3r$) and $s<-a$  (since then $s-\h\le s+a<0$ for all $\3r$). If $s>a$, $g(\t-\z)$ peaks when $\h=a$ (the positive $z$ axis $\5Z_+$), and if $s<-a$, it peaks when $\h=-a$ (the negative $z$ axis $\5Z_-$). Hence $g(\t-\z)$ is a \sl beam-shaping pulse function, \rm forming a beam along $\5Z_+$ or $\5Z_-$ (depending on the choice of $s$) through its dependence on $\z$.  Such pulses are the basis for \sl complex-source pulsed-beams \rm  in the engineering literature; see \ci{HF89} and the detailed references given in \ci{HK9}.

\section{Electric-magnetic dipoles and quasar engines}\label{EM dipoles}

Equations \eq{rels} also have solutions representing \sl non-axially symmetric \rm generalizations of \eq{F-}. The simplest ones are
\begin{align}\label{Fpm}
\3F_+=\b(u,v,\y_\pm)\3\c_+  \qqq \3F_-=\g(u,v,\y_\pm)\3\c_-
\end{align}
where \eq{rels}  require that
\begin{align*}
\b_u=\b_-=0\ \ \hbox{and}\ \ \g_v=\g_+=0,
\end{align*}
respectively.
Thus $\3F_+$ must be \sl advanced (absorbed) \rm and independent of $\y_-$, and $\3F_-$ must be \sl retarded (emitted) \rm and independent of $\y_+$:
\begin{align}\label{F2}
\3F_+=\b(\t+\z,\y_+)\3\c_+\ \ \hbox{and}\ \  \3F_-=\g(\t-\z,\y_-)\3\c_-
\end{align}
where the dependence on $\f$ through $\y_\pm$ must be continuous and periodic. It is easy to verify directly that $\3F_\pm$  satisfy the homogeneous Maxwell equations outside $\5Q$, so their charge-current densities must be supported on $\5Q$.

The fields \eq{F2} are \sl null: \rm 
\begin{align}\label{rad}
\3F_\pm^2\equiv (\3E_\pm+i\3B_\pm)^2=0, \ \ \hbox{\ie}\ \  
\3E_\pm^2-\3B_\pm^2=\3E_\pm\cdot\3B_\pm=0.
\end{align}
These are the conditions of \sl pure radiation \rm everywhere outside $\5Q$. As proved in \ci{K11, K11b}, this means that $\3F_\pm$ propagate without leaving any \sl reactive (rest) energy \rm behind; all their energy flows at the speed of light $c$, which is not true of generic EM fields, even in vacuum. Normally, electromagnetic energy flows at speeds $v<c$ in the near zone due, among other things, to interference between field components propagating in different directions.

Thus $\3F_\pm$ are \sl extreme. \rm  To obtain `normal' solutions whose energies flow at speeds less than $c$ in the near zone, we must include a longitudinal component $\a(u,v,\y_\pm)\ne 0$. This makes \eq{rels} more difficult to solve and will be pursued elsewhere. 

By \eq{cpm}, $\3F_\pm$ satisfies 
\begin{align}\label{hel}
\bh\z\times\3F_\pm=\pm i\3F_\pm,
\end{align}
and we have seen that $\3F_\pm$ depends on time only through $\t\pm\z$. We have interpreted this by saying that $\3F_+$ is \sl absorbed \rm and $\3F_-$ is \sl emitted. \rm  To confirm this,  consider the limit of \eq{cpm} as $a\to 0$, where $\bh\z\to\bh r$:
\begin{align}\label{hel2}
\bh r\times\3F_\pm=\pm i\3F_\pm\iff \bh r\times\3E_\pm=\mp\3B_\pm\ \ \hbox{and}\ \  \bh r\times\3B_\pm=\pm\3E_\pm.
\end{align}
This gives the Poynting vectors
\begin{align}\label{hel3}
\3S_\pm\equiv \3E_\pm\times\3B_\pm=\mp\3E_\pm\times(\bh r\times\3E_\pm)=\mp\3E_\pm^2\,\bh r,
\end{align}
so the energy of $\3F_+$  flows \sl inward \rm to the origin and that of $\3F_-$ flows \sl outward \rm from the origin. This is indeed consistent with $\3F_+$ being absorbed and $\3F_-$ being emitted. 

Let us derive the spheroidal version of \eq{hel3}. It follows from \eq{hel} that the Poynting vector is given by
\begin{align*}
\mp 2\3S_\pm=\pm i\3F_\pm^*\times\3F_\pm=\3F_\pm^*\times(\bh\z\times\3F_\pm)
=|\3F_\pm|^2\bh\z-(\bh\z\cdot\3F_\pm^*)\3F_\pm,
\end{align*}
hence
\begin{align}\label{S}
\mp\3S_\pm=U_\pm\bh\z-\frac12(\bh\z\cdot\3F_\pm^*)\3F_\pm
\end{align}
where
\begin{align*}
U_\pm=\frac12|\3F_\pm|^2=\frac12(\3E_\pm^2+\3B_\pm^2)=\3E_\pm^2
\end{align*}
is the energy density. The left side of \eq{S} is real while the right side is complex. Since $\bh\z^*\cdot\3F_\pm^*=0$ and $\bh\z+\bh\z^*=2\grad\x$, \eq{S} can be rewritten as
\begin{gather*}
\mp\3S_\pm=U_\pm\bh\z-(\grad\x\cdot\3F_\pm^*)\3F_\pm=U_\pm\bh\z-\grad\x\cdot(\3E_\pm-i\3B_\pm)(\3E_\pm+i\3B_\pm).
\end{gather*}
Taking the real part of both sides gives
\begin{align}\label{S1}
\mp\3S_\pm=U_\pm\grad\x-\grad\x\cdot\3E_\pm\3E_\pm-\grad\x\cdot\3B_\pm\3B_\pm=-\grad\x\cdot\4T_\pm
\end{align}
where $\4T_\pm$ is the Maxwell stress tensor of $\3F_\pm$. This is the spheroidal version of \eq{hel3}, reducing to the latter as $a\to0$ since $\grad\x\to\bh r$ and $\grad\x\cdot\3E_\pm, \grad\x\cdot\3B_\pm$ both vanish. 

We now introduce the angular functions
\begin{align}\label{Zm}
Z^m_\pm&\equiv e^{-im\y_\pm}= e^{-im\f}\tan^{\pm m}(\vq/2)=e^{-im\f}\lp\frac{1\mp\cos\vq}{1\pm\cos\vq}\rp^{m/2}\\
&\qq =e^{-im\f}\lp\frac{\z\mp\2z}{\z\pm\2z}\rp^{m/2} \ \ \hbox{with}\ \  m\in\4Z,\nt
\end{align}
where $m$ must be an integer for $Z^m_\pm$ to be single-valued. If $\pm m>0$, $\tan^{\pm m}(\vq/2)$ is a  \sl vortex factor \rm suppressing the positive $z$ axis and amplifying the negative $z$ axis; if $\pm m<0$, it does the reverse. This can be seen by noting that $z=\x\h/a$ \eq{zeta2}, hence
\begin{align*}
\z\pm\2z=(\x-i\h)\pm (z-ia)=(\x\pm z)\mp i(a\pm\h)=\frac{\x\mp ia}a(a\pm \h)
\end{align*}
and
\begin{align*}
\z\pm \2z=0\iff \h=\mp a\iff \r=0,\ \mp z\ge 0.
\end{align*}
Define the \sl basic null helicity wavelets with angular momentum $m$ about the $z$ axis \rm  by
\begin{align}\label{Fmpm}
\bx{\3F^m_\pm=g'(\t\pm\z)Z^m_\pm\3\c_\pm\,.}
\end{align}
They form a Fourier basis for absorbed and emitted helicity fields analytic in $\y_\pm$ and periodic in $\f$.

To explain and justify the term \sl helicity wavelets, \rm start with the time-harmonic case.
From \eq{ast} it follows that 
\begin{align}\label{ast1}
g_0\0t=e^{-i\o t}\imp g(t-is)=\sgn s\cdot H(\o s) e^{-i\o(t-is)},
\end{align}
where $H$ is the Heaviside function (see also  \ci[page R298]{K3}). Hence
\begin{align*}
g'(\t\pm\z)=-i\o \sgn(s\pm\h) H(\o(s\pm\h)) e^{-i\o(t\pm\x)} e^{-\o(s\pm\h)}.
\end{align*}
At any fixed position $\3r\notin\5Q$,
\begin{align}\label{helicity}
\3F^m_\pm=Ae^{-i\o t}\3\c_\pm=\frac A\r\,e^{-i\o t}(\bh\f\pm i\bh\vq),
\end{align}
where $A$ is a complex function of $\3r, \3a, s, m$ and $\o$. If $s>0$, then $\o>0$ by \eq{ast1} and \eq{helicity} shows that the polarization of $\3F^m_-$ spins clockwise about the vector $\3r$, as viewed from the origin, and that of $\3F^m_+$ spins counterclockwise. Since $\3F_-$ is retarded and $\3F_+$ is advanced, both wavelets have negative helicity for $\o>0$; that is, the polarization of $\3F_+$ spins clockwise when viewed in the direction of propagation, \sl toward \rm the origin. If $\o<0$, these spins are reversed. Thus we have shown:

\sl For time-harmonic signals, the wavelets $\3F^m_\pm$ have helicity\,\footnote{In \ci{K11} and \ci{K4a}, I used the opposite convention, so the helicity there was $\sgn\o$. This can be arranged, for example, by defining $\t=t+is$.
}
$-\sgn\o$. \rm

Furthermore, the above concept of helicity makes sense even when $g_0$ is \sl not \rm time-harmonic. From \eq{ast1} it follows that for a general signal, $g\0\t$ consists of the positive-frequency part of $g_0$ if $s>0$ and the negative-frequency part of $g_0$ if $s<0$:
\begin{align*}
s>0&\imp g\0\t=\frac1{2\p}\int_0^\infty\dd\o\,e^{-i\o t-\o s}\1g_0\0\o\equiv g_0^+\0\t\\
s<0&\imp g\0\t=-\frac1{2\p}\int_{-\infty}^0\dd\o\,e^{-i\o t-\o s}\1g_0\0\o\equiv -g_0^-\0\t,
\end{align*}
where $\1g_0\0\o$ is the ordinary Fourier transform of $g_0\0t$.

\begin{itemize}
\item \sl When $s>a$, $\3F_\pm$ is a superposition of negative-helicity fields.
\item When $s<-a$,  $\3F_\pm$ is a superposition of positive-helicity fields. \rm 
\end{itemize}

This idea extends to fields of indefinite helicity as follows. Define the \sl helicity operator \rm $\5S$ on solutions as multiplication by $-\sgn\o$ in the Fourier domain. This can be expressed directly in the spacetime domain as $-i$ times the \sl temporal Hilbert transform: \rm 
\begin{align}\label{H}
\5S\3F\equiv \frac1{i\p}\ppint_{-\infty}^\infty\frac{\dd t'}{t'-t}\3F(\3r, t')=\3F^-\rt-\3F^+\rt,
\end{align}
where $\ppint$ denotes the principal-value integral and $\3F^\pm$ denote the positive- and negative frequency parts of $\3F$.

\bf Remark 1. \rm It may be thought unphysical to restrict the time-harmonic field \eq{helicity} to just positive or just negative frequencies. This would be justified if the opposite component can be obtained by complex conjugation using a reality condition.\footnote{This is commonly done in the engineering literature, with the understanding that the true field is obtained by thing the real part.
}
However, since $\3F$ is complex, its positive- and negative-frequency components are \sl independent. \rm The reality conditions
\begin{align}\label{reality}
\3E_\o\orr^*=\3E_{-\o}\orr, \qq \3B_\o\orr^*=\3B_{-\o}\orr
\end{align}
apply only to the Fourier components of the real fields $\3E, \3B$. A \sl general \rm time-harmonic field has the complex form
\begin{align*}
\3F\rt=\3F_\o\orr e^{-i\o t}+\3F_{-\o}\orr e^{i\o t}=\3E\rt+i\3B\rt,\qq \o>0,
\end{align*}
where $\3F_\o$ and $\3F_{-\o}$ can be chosen independently. Then the real fields are
\begin{align*}
\3E&=\re\3F=\frac12[\3F_\o+\3F_{-\o}^*]e^{-i\o t}+\frac12[\3F_{-\o}+\3F_\o^*]e^{i\o t}\\
\3B&=\im\3F=\frac1{2i}[\3F_\o-\3F_{-\o}^*]e^{-i\o t}+\frac1{2i}[\3F_{-\o}-\3F_\o^*]e^{i\o t},
\end{align*}
so
\begin{align*}
\3E_\o&=\frac12[\3F_\o+\3F_{-\o}^*] =\3E_{-\o}^*&&
\3B_\o=\frac1{2i}[\3F_\o-\3F_{-\o}^*] =\3B_{-\o}^*
\end{align*}
as required. As we have seen, $\3F_\o\orr e^{-i\o t}$ gives a real field of negative helicity and $\3F_{-\o}\orr e^{i\o t}$  gives a real field of positive helicity. See \ci{K4a}.

\bf Remark 2. \rm Electromagnetic wavelets of definite helicity were already constructed in 1994 \ci{K11a}. However, those wavelets were \sl globally sourceless, \rm hence unrealizable.

\section{Regularized charge-current density of $\3F_\pm$}\label{S:reg}

Let us write the general helicity fields \eq{F2} as
\begin{align}\label{F3}
\3F_\s\rt=g'(\t+\s\z,\y_\s)\3\c_\s,\qqq\s=\pm 1.
\end{align}
They satisfy the homogeneous Maxwell equations everywhere outside the singularity set $\5Q$:
\begin{align}\label{Max0}
\3r\notin\5Q\imp \div\3F=0\ \ \hbox{and}\ \  \pl_t\3F_\s+i\curl\3F_\s=\30,
\end{align}
hence their charge-current density $(\vr_\s, \3J_\s)$ must be a distribution supported on $\5Q$. We shall now compute this distribution by \sl regularizing \rm the field $\3F_\s$ and then applying Maxwell's equations `in reverse' to compute the sources. Let $\e>0,\  0<b<a$, choose differentiable functions $e\0\x$ and $h_\pm\0\h$ with
\begin{align}\label{eh}
 e\0\x=\begin{cases}
0,&\x\le 0\\ 1,&\x>\e,
\end{cases}&&
 h_\pm\0\h=\begin{cases}
1,&\h< b\\ 0,&\h\ge a,
\end{cases}
\end{align}
and let
\begin{align}\label{R}
R(\x,\h)=e\0\x h_+\0\h h_-(-\h).
\end{align}
Then
\begin{align*}
R(0,\h)=R(\x,\pm a)=0, \ \ \hbox{hence}\ \ \3r\in\5Q\imp R=0.
\end{align*}
Furthermore, define the \sl solid \rm spheroid $\ctil E^\e$ and semi-hyperboloids $\ctil H^b_\pm$
\begin{gather}\label{EeHb}
\ctil E^\e=\{\3r: 0\le \x\le\e\} \qqq \ctil H^b_\pm=\{\3r: b\le \pm\h\le a\}
\end{gather}
and their union
\begin{align}\label{Qeb}
\ctil Q^{\e,b}=\ctil E^\e\cup\ctil H^b_+\cup\ctil H^b_-.
\end{align}
Then
\begin{align*}
\3r\notin\ctil Q^{\e,b}\imp R=1.
\end{align*}
As $\e\to0$ and $b\to a$, $\ctil Q^{\e,b}\to\5Q$. We shall regularize $\3F_\s$ by multiplying it by $R$, thus eliminating the singularity on $\5Q$ while leaving the field unchanged outside of $\ctil Q^{\e,b}$.  We compute the charge-current density of the regularized field and then find the `bare' charge-current density of $\3F_\s$ as its limit.

Thus define the \sl regularized field \rm
\begin{align}\label{2Fs}
\bt F_\s=R\3F_\s.
\end{align}
Taking the divergence \sl formally, \rm we obtain the charge density of $\3F_\s$ as
\begin{align*}
\2\vr_\s\equiv \div\bt F_\s&=R\,\div\3F_\s+\pl_\x R\,\grad\x\cdot\3F_\s+\pl_\h R\,\grad\h\cdot\3F_\s\\
&=R\,\div\3F_\s+\frac{\s a}{\x^2+\h^2}\,g'(\t+\s\z,\y_\s)(\pl_\x-i\pl_\h) R,
\end{align*}
where we have used \eq{xdotchi} and \eq{hdotchi} from the Appendix.
Assume that as $\3r\to\5Q$, $R$ vanishes sufficiently rapidly that the first term is identically zero. (This simply requires some smoothness of $R$ at $\x=0$ and $\h=\pm a$.)
Then
\begin{align}\label{2r2}
\2\vr_\s=\frac{2\s a}{\x^2+\h^2}\,g'(\t+\s\z,\y_\s)\pl_\z^* R\,, 
\end{align}
where $\pl_\z^*=\tfrac12(\pl_\x-i\pl_\h)$ is the partial derivative with respect to the complex conjugate $\z^*$. In terms of $e$ nd $h_\pm$ we have
\begin{align}\label{2r3}
\2\vr_\s=\2\vr_\s^e+\2\vr_\s^++\2\vr_\s^-,
\end{align}
where
\begin{align}\label{2r31}
\2\vr_\s^e&=e'\0\x h_+\0\h h_-(-\h)\grad\x\cdot\3F_\s\\
\2\vr_\s^\pm&=\pm e\0\x h'_\pm(\pm\h)\grad\h\cdot\3F_\s \nt
\end{align}
and we have used the fact that
\begin{align*}
h_+'(\h)\ne 0&\imp b<\h<a\imp h_-(-\h)=1\\
h_-'(-\h)\ne 0&\imp -a<\h<-b\imp h_+(\h)=1.
\end{align*}
Similarly, we compute the current density of $\bt F_\s$ as
\begin{align*}
-\bt J_\s&\equiv \pl_t\bt F_\s+i\curl\bt F_\s\\
&=R\lp\pl_t\3F_\s+i\curl\3F_\s\rp+i\pl_\x R\,\grad\x\times\bt F_\s+i\pl_\h R\,\grad\h\times\bt F_\s.
\end{align*}
Again, the first term vanishes and we have
\begin{align}\label{2J3}
\bt J_\s&=-i\pl_\x R\grad\x\times\bt F_\s-i\pl_\h R\grad\h\times\bt F_\s\\
&=\bt J_\s^e+\bt J_\s^++\bt J_\s^-\nt
\end{align}
where
\begin{align}\label{2J31}
\bt J_\s^e&=-ie'\0\x h_+\0\h h_-(-\h)\grad\x\times\3F_\s\\
\bt J_\s^\pm&=\mp ie\0\x h'_\pm(\pm\h)\grad\h\times\3F_\s.\nt
\end{align}
Since the supports of $e'\0\x$ and $h'_\pm\0\h$ are $[0,\e]$ and $[b,a]$, respectively, 
it follows that $(\2\vr_\s^e,\bt J_\s^e)$ are supported in $\ctil E^\e$
and $(\2\vr_\s^\pm,\bt J_\s^\pm)$ are supported in $\ctil H^b_\pm$. 

The regularized currents can be expressed compactly by defining
\begin{align}\label{2v3}
\bt v_\s^e&\equiv -i\frac{\grad\x\times\3\c_\s}{\grad\x\cdot\3\c_\s}=
-\frac{i\s\x}a\bh z+\frac{i\s\h}a\sr{\frac{a^2+\x^2}{a^2-\h^2}}\,\bh\r+\sr{\frac{a^2+\x^2}{a^2-\h^2}}\,\bh\f\\
\bt v_\s^\pm&\equiv -i\frac{\grad\h\times\3\c_\s}{\grad\h\cdot\3\c_\s}=
-\frac{\s\h}a\bh z-\frac{\s\x}a\sr{\frac{a^2-\h^2}{a^2+\x^2}}\,\bh\r+\sr{\frac{a^2-\h^2}{a^2+\x^2}}\,\bh\f,\nt
\end{align}
where we have used the identity $a\r=\sr{m_\x m_\h}$ \eq{zeta2}. According to \eq{2r31} and \eq{2J31}, the currents are given by\footnote{Although $\bt v_\s^+=\bt v_\s^-$, recall that $\h$ in $\2\vr_\s^\pm$ and $\bt J_\s^\pm$ is restricted to $b\le\pm\h\le a$.
}
\begin{align}\label{Jvr}
\bt J_\s^e=\bt v_\s^e\2\vr_\s^e\ \ \hbox{and}\ \ \bt J_\s^\pm=\bt v_\s^\pm\2\vr_\s^\pm.
\end{align}
Note that
\begin{align}\label{v2=1}
\bt v_\s^e\cdot\bt v_\s^e=1\ \ \hbox{and}\ \ \bt v_\s^\pm\cdot\bt v_\s^\pm=1,
\end{align}
the significance of which will be investigated elsewhere.\footnote{It is tempting to interpret $\bt v_\s^e$ and $ \bt v_\s^\pm$ as `charge-flow velocities,' in which case \eq{v2=1} would state that the charges on $\ctil E^\e$ and $\ctil H^b_\pm$ flow at the speed of light $c\equiv 1$. However, such an interpretation is valid only if the associated densities have a definite sign. For example, equal and opposite uniform charge distributions flowing in equal and opposite directions give a vanishing charge density but double the current. This suggests a `two-fluid' interpretation whereby the sources are split into ones with positive and negative charges. (A further challenge is presented by the fact that $\bt v_\s^e$ is complex.) 
}

Equations \eq{2v3} contain a wealth of information. For example,  $\bh\f\cdot\bt v_\s^e$ becomes infinite as $\3r\to\5Z_\pm\,(\h\to \pm a$), showing that $\ctil E^\e$ has  \sl vortices \rm there.  Since $\bt v_\s^\pm$ is real, it relates the electric current density to the electric charge density and the magnetic current density to the magnetic charge density. On the other hand, the $z$ and $\r$ components of $\bt v_\s^e$ are imaginary, so they convert electric to magnetic sources and vice versa. 

The \sl unregularized \rm or `bare' sources are obtained by taking the limits $\e\to0$ and $b\to a$. By \eq{eh},
\begin{align*}
\e\to0\imp e\0\x\to H\0\x\ \ \hbox{and}\ \  b\to a\imp h_\pm\0\h\to H(a-\h),
\end{align*}
where $H$ is the Heaviside step function. Therefore
\begin{align*}
e'\0\x\to\d\0\x,\qq h_+'\0\h\to-\d(a-\h),\qq -h_-'(-\h)\to \d(a+\h)
\end{align*}
restrict the sources $(\vr_\s, \3J_\s)$ to $\5D$ and $\5Z_\pm$, confirming that the bare sources are indeed distributions supported on these sets. Equations \eq{2r31} give, after some simplification,
\begin{align*}
\vr_\s^e&=\d\0\x\frac{\s a}{\h^2} g'(\t-i\s\h,\y_\s)\\
\vr_\s^\pm&=\pm i\d(a\mp\h)\frac{\s a}{\x^2+a^2} g'(\t+\s\x\mp i\d a,\y_\s),
\end{align*}
where we have dropped the superfluous factors $H\0\x$ and $H(a\mp\h)$.

The unregularized versions of $\bt v_\s^e$ and $\bt v_\s^\pm$ are
\begin{align*}
\3v_\s^e=\frac{i\s\h\bh\r+a\bh\f}\r \ \ \hbox{and}\ \  
\3v_\s^\pm=\mp\s\bh z,
\end{align*}
where we have used the fact that $\sr{a^2-\h^2}=\r$ on $\5D$.

\section{The total charge vanishes: no monopoles}\label{Q=0}

As noted above, the regualrized sources $\2\vr_\s, \bt J_\s$ are \sl complex  \rm and their real and imaginary parts are the electric and magnetic charge-current densities for $\bt F_\s$. The total charge $\2Q\0\t$ must be constant due to charge conservation. But if this constant were nonzero, it would have to be \sl real \rm to avoid magnetic monopoles, and this is impossible due to analyticity in $\t$, unless $\2Q\0\t\equiv 0$.  In this section we prove that $\2Q\0\t$ does indeed vanish identically. It suffices to prove this for the regularized basic helicity wavelets
\begin{align}\label{2Fm}
\bt F_\s^m&=R(\x,\h)\3F_\s^m=R(\x,\h)g'(\t+\s\z)e^{-im\y_\s}\3\c_\s.
\end{align}
The volume element in oblate spheroidal coordinates is given by  \ci{K3}
\begin{align*}
\dd^3\3r=\frac{\x^2+\h^2}a\,\dd\x\,\dd\h\,\dd\f,
\end{align*}
and
\begin{align*}
\int_0^{2\p}\dd\f\, e^{-im\y_\s}=\int_0^{2\p}\dd\f\, e^{-im\f\pm m\k}=2\p\d_m^0
\end{align*}
since $\k=\ln\tan(\vq/2)$ is independent of $\f$.
Equations \eq{2r2} thus give the total charge of $\bt F^m_\s$ as
\begin{align*}
\2Q^m_\s\0\t&=2\p\s\d_m^0\int_0^\infty\dd\x\int_{-a}^a\dd\h\, g'(\t+\s\x-i\s\h)
\lb \pl_\x R-i\pl_\h R\rb.
\end{align*}
Integrating the first term by parts in $\x$ and the second term in $\h$, we obtain
\begin{align*}
\2Q^m_\s\0\t&=2\p\s\d_m^0\int_{-a}^a\dd\h\,\lb R g'\rb_{\x=0}^\infty
-2\p i\s\d_m^0\int_0^\infty\dd\x\,\lb R g'\rb_{\h=-a}^a\\
&\qq -4\p\s\d_m^0\int_0^\infty\dd\x\int_{-a}^a\dd\h\,R\, \pl_\z^* g'(\t+\s\z).
\end{align*}
The first boundary term vanishes because $R(0,\h)=0$ and 
\begin{align*}
\x\to\infty\imp g'(\t+\s\x-i\s\h)\to 0.
\end{align*}
The second boundary term vanishes because $R(\x, \pm a)=0$. Finally, the last term vanishes because $g'(\t+\s\z)$ is \sl analytic \rm in $\z=\x-i\h$. This proves that
\begin{align}\label{2Q0}
\2Q^m_\s\0\t\equiv 0
\end{align}
as claimed. Thus \sl $\bt F^m_\s$ has neither magnetic nor electric monopoles. \rm

Since the total charge vanishes, the fields $\bt F^m_\s$ are due to a combination of electric and magnetic \sl polarizations. \rm Hence the support $\ctil Q^{\e,b}$ \eq{Qeb} of the charge-current density acts as a combined \sl electric-magnetic dipole antenna for receiving $\bt F^m_+$ or emitting $\bt F^m_-$. \rm  If $\ctil E^\e$ is \sl flat \rm and $\ctil H^b_\pm$  are \sl narrow, \rm meaning that
\begin{align*}
\frac\e a\ll 1\ \ \hbox{and}\ \ \frac{a-b}a\ll 1,
\end{align*}
then the electric dipole is supported mostly on $\ctil H^b_+\cup\ctil H^b_-$ and the magnetic dipole is supported mostly on $\ctil E^\e$. 

\begin{figure}[th]
\sh8\includegraphics[width=4.5 in]{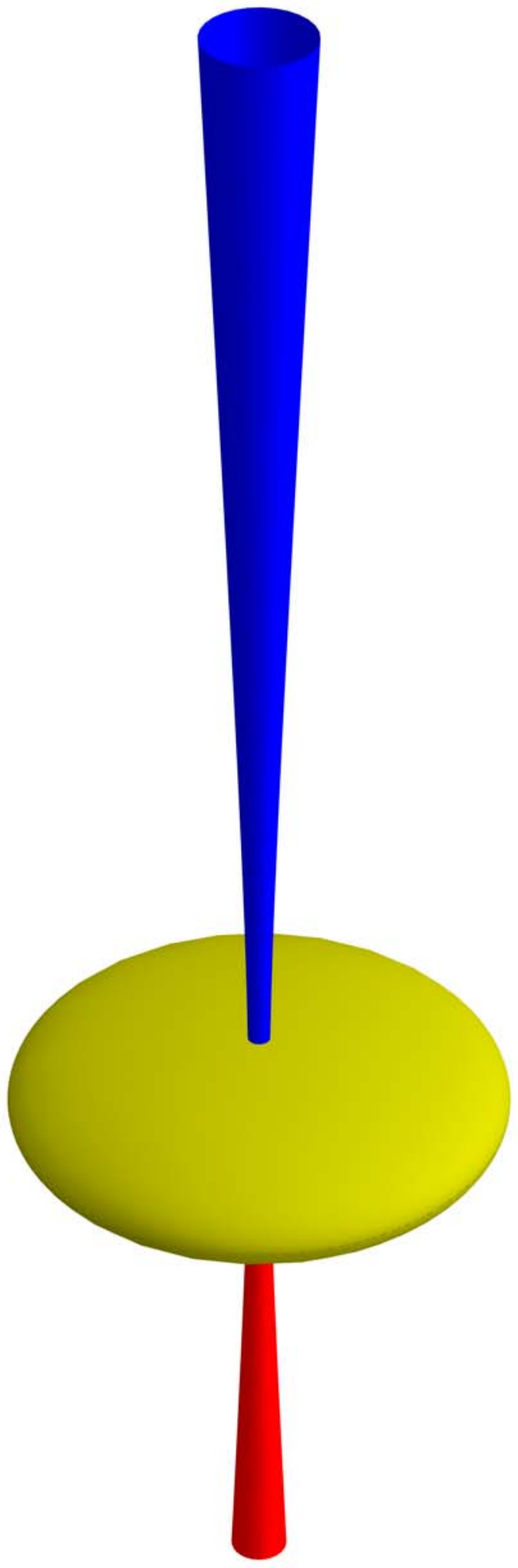}
\caption{\small The charge-current densities \eq{2r31}, \eq{2J31} of the regularized fields $\bt F_\s$ \eq{2Fs} are distributions supported on the union $\ctil Q^{\e,b}$ \eq{Qeb} of a solid oblate spheroid $\ctil E^\e$ and two solid semi-hyperboloids $\ctil H_\pm^b$ \eq{EeHb}.  According to \eq{2Q0}, the total electric and magnetic charges vanish, hence $\ctil Q^{\e,b}$ acts as an \sl electric-magnetic dipole antenna. \rm   $\5H^\e_\pm$ support the electric dipole and $\ctil E^\e$ supports the magnetic dipole, and the flow of electric and magnetic currents within and between $\ctil H_\pm^b$ and $\ctil E^\e$ is determined by the analytic signal $g\0\t$ \eq{ast}. In the `bare' (unregularized) limit ($\e\to0, b\to a$), $\ctil E^b$ approaches the branch disk $\5D$ and $\ctil H_\pm^b$ approach $\5Z_\pm$, so $\ctil Q^{\e,b}\to\5Q$ \eq{5Q}. As explained in the text, $\ctil Q^{\e,b}$ could be used to model \sl quasar engines, \rm with  $\ctil E^\e$ and $\ctil H_\pm^b$ representing the accretion disk and the jets, respectively.}
\label{F:EMDipole}
\end{figure}

We propose this system as an electromagnetic model for \sl quasar engines,\rm\footnote{We are obviously ignoring gravity by working in flat spacetime. Our model has roughly the same relation to quasars as Newman's holomorphic Coulomb field \ci{N73, K4} has to spinning (Kerr-Newman) black holes.
}
with $\ctil E^\e$ representing the \sl accretion disk \rm and $\ctil H_\pm^b$ the \sl vortex jets; \rm see Figure \ref{F:EMDipole}. It is known that quasars radiate light with a high degree of helicity (circular polarization) \ci{BF2, E3}.  Furthermore, since quasars are the most distant visible objects in the universe, they must radiate extremely powerful and highly collimated beams. The radiated wavelets $\3F^m_-$ in \eq{Fmpm} and their regularized versions $\bt F^m_-$ derive their power and collimation from a combination of the beam-shaping properties of $g'(\t-\z)$, their angular momentum, and the associated vortex factor $[(\z+\2z)/(\z-\2z)]^{m/2}$ in \eq{Zm}.

\bf Remark. \rm The above is a highly idealized and simplified model for quasars. Among other things, it ignores the fact that the accretion disk and the jets of real quasars consist of \sl plasmas \rm and the jets are coupled to the differentially rotating accretion disk by magnetohydrodynamic equations, or preferably a relativistic version thereof. Note that the shape of $e\0\x$ is related to the density of $\ctil E^\e$, and the shapes of $h_\pm\0\h$ are related to the densities of $\ctil H_\pm^b$. If the plasma dynamics can be expressed in our spheroidal coordinate system, as suggested by Professor En\ss lin (private communication), then the coupling between the accretion disk and the jets could perhaps be represented by a relation between $e\0\x$ and $h_\pm(\pm\h)$. This would be an interesting subject for future study.

\section*{Appendix}\label{S:appendix}

Here we derive various expressions needed for the above computations. From the definition \eq{zeta} and \eq{vq} we have
\begin{gather}\label{zeta2}
\z^2=r^2-a^2-2iaz=(\x-i\h)^2\imp\x^2-\h^2=r^2-a^2,\qq az=\x\h\\
a^2\r^2=a^2r^2-a^2z^2=a^2(\x^2-\h^2)+a^4-\x^2\h^2=m_\x m_\h \nt
\end{gather}
where
\begin{align*}
m_\x =a^2+\x^2,\qq m_\h=a^2-\h^2.
\end{align*}
\begin{align*}
\bh\z&=\grad\x-i\grad\h=\frac{\3r-i\3a}\z=\frac{\bh\r\r+\bh z\2z}\z=\bh\r\sin\vq+\bh z\cos\vq&&\bh\z^2=1\\
\bh\vq&=\bh\f\times\bh\z=\bh z\sin\vq-\bh\r\cos\vq&&\bh\vq^2=1.
\end{align*}
Taking the real and imaginary parts of $\bh\z$ and simplifying gives
\begin{align*}
\grad\x&=\frac{\x\3r+\h\3a}\m=\frac{\x\r}\m\bh\r+\frac{\h n_\x}a\bh z,
\ \ \hbox{where}\ \  n_\x=\frac{m_\x}\m,\qq \m=|\x|^2\\
\grad\h&=\frac{\x\3a-\h\3r}\m=\frac{\x n_\h}a\bh z-\frac{\h\r}\m\,\bh\r,\ \ \hbox{where}\ \ n_\h=\frac{m_\h}\m.
\end{align*}
It follows that
\begin{align*}
|\grad\x|^2=n_\x\ \ \hbox{and}\ \  |\grad\h|^2=n_\h.
\end{align*}
To compute the regularized sources in Section \ref{S:reg}, we need expressions for $\grad\x\cdot\3\c_\s, \grad\h\cdot\3\c_\s$ and $\grad\x\times\3\c_\s, \grad\h\times\3\c_\s$, where $\s=\pm$. It follows that
\begin{align}\label{xxh}
\grad\x\times\grad\h=-\frac{a\r}\m\bh\f,
\end{align}
and since $\grad\x\cdot\bh\f=0$, we have
\begin{align}\label{xdotchi}
\grad\x\cdot\3\c_\s&=i\s\grad\x\cdot\frac{\bh\vq}\r
=i\s\grad\x\cdot\frac{\bh\f\times\bh\z}\r=-i\s\bh\f\cdot\frac{\grad\x\times\bh\z}\r\\
&=-\s\bh\f\cdot\frac{\grad\x\times\grad\h}\r=\s\frac{a}\m. \nt
\end{align}
Furthermore,
\begin{align}\label{hdotchi}
\bh\z\cdot\3\c_\s=(\grad\x-i\grad\h)\cdot\3\c_\s=0\imp\grad\h\cdot\3\c_\s=-i\s\frac{a}\m.
\end{align}
Equations \eq{xdotchi} and \eq{hdotchi} will be used in Section \ref{S:reg} to compute the regularized charge densities. To compute the regularized currents, we need the following:
\begin{align*}
\grad\x\times\bh\f&=\frac{\x\r}\m \bh z\,-\frac{\h n_\x}a\bh\r\,\\
\grad\x\times\bh\vq&=\grad\x\times(\bh\f\times\bh\z)=|\grad\x|^2\bh\f=n_\x\bh\f\\
\grad\h\times\bh\f&=-\frac{\h\r}\m \bh z-\frac{\x n_\h}a\bh\r\\
\grad\h\times\bh\vq&=\grad\h\times(\bh\f\times\bh\z)=-i|\grad\h|^2\bh\f\,=-in_\h \bh\f,
\end{align*}
hence
\begin{align}\label{timeschi}
\grad\x\times\3\c_\s&=\frac\x\m\bh z-\frac{\h n_\x}{a\r}\bh\r+i\s\frac{n_\x}\r\bh\f\\
\grad\h\times\3\c_\s&=-\frac\h\m\bh z-\frac{\x n_\h}{a\r}\bh\r+\s\frac{n_\h}\r\bh\f. \nt
\end{align}

\section*{Acknowledgements}
This work was supported by AFOSR Grant \#FA9550-12-1-0122. 
I thank Sir Roger Penrose for suggesting, at a conference in 2000, that my methods could be useful for modeling quasars. I also thank Professor Torsten En\ss ling for pointing out the need to include the study of magnetohydrodynamic coupling between the accretion disk and the jets of quasars.

\end{document}